\pdfoutput=1
\newcommand*{\ATLASLATEXPATH}{latex/}
\documentclass[cernpreprint, texlive=2016, UKenglish]{\ATLASLATEXPATH atlasdoc}
\usepackage{\ATLASLATEXPATH atlaspackage}
\usepackage{\ATLASLATEXPATH atlasbiblatex}
\usepackage[hion=true,process=true]{\ATLASLATEXPATH atlasphysics}
\graphicspath{{logos/}{figures/}}
\AtlasTitle{Prompt and non-prompt $J/\psi$ elliptic flow in Pb+Pb collisions at $\sqrt{s_{_\text{NN}}} = 5.02$~\TeV\ with the ATLAS detector}
\AtlasAbstract{%
The elliptic flow of prompt and non-prompt $J/\psi$ was measured in the dimuon decay channel in Pb+Pb collisions at $\sqrt{s_{_\text{NN}}}=5.02$~\TeV\ with an integrated luminosity of $0.42~\mathrm{nb}^{-1}$ with the ATLAS detector at the LHC. The prompt and non-prompt signals are separated using a two-dimensional simultaneous fit of the invariant mass and pseudo-proper decay time of the dimuon system from the $J/\psi$ decay. The measurement is performed in the kinematic range of dimuon transverse momentum and rapidity $9<p_\mathrm{T}<30$~\GeV, $|y|<2$, and 0--60\% collision centrality. The elliptic flow coefficient, $v_2$, is evaluated relative to the event plane and the results are presented as a function of transverse momentum, rapidity and centrality. It is found that prompt and non-prompt $J/\psi$ mesons have non-zero elliptic flow. Prompt $J/\psi$ $v_2$ decreases as a function of $p_\mathrm{T}$, while for non-prompt $J/\psi$ it is, with limited statistical significance, consistent with a flat behaviour over the studied kinematic region. There is no observed dependence on rapidity or centrality.
}

\AtlasRefCode{HION-2017-05}

\PreprintIdNumber{CERN-EP-2018-170}

\arXivId{1807.05198}

\AtlasJournal{Eur. Phys. J. C}
\AtlasJournalRef{\EPJC 78 (2018) 784}
\AtlasDOI{10.1140/epjc/s10052-018-6243-9}

\hypersetup{pdftitle={ATLAS document},pdfauthor={The ATLAS Collaboration}}
\begin{document}
\maketitle
\section{Introduction}
With the advent of lead--lead collisions at the centre-of-mass energy of $5.02~\TeV$ per nucleon--nucleon pair, new opportunities open up for understanding the detailed properties of the hot dense plasma produced in such collisions. A special advantage of studies with quarkonia as hard probes of the plasma properties is that the comparison of prompt and non-prompt \jpsi mesons elucidates the differences between the responses of the produced $c$-quark and $b$-quark systems. This is because the prompt \jpsi mesons are \ccbar systems produced soon after the collision whereas the non-prompt \jpsi mesons come from decays of $b$-hadrons that are formed outside the medium~\cite{Kopeliovich:2017imm}. Thus, the comparison of these two classes of \jpsi mesons probes the flavour dependence of the mechanisms of the interactions of heavy quarks with the medium. ATLAS measurements of the attenuation of both the prompt and non-prompt \jpsi meson yields indicate very strong medium effects that are surprisingly similar in magnitude at this collision energy~\cite{Tapia:2018quy}.

A complementary and powerful probe into the heavy-quark flavour dependence of interaction mechanisms can be obtained by studying the azimuthal asymmetries of prompt and non-prompt quarkonia. Such studies~\cite{ALICE2013,ALICE2017,CMS2017} are especially useful in the dimuon transverse momentum range $9<\pT<30~\GeV$, investigated in this paper, since this range represents the transition between the lower \pT region, in which recombination processes are believed to play an important role~\cite{Zhou:2014kka,Yan:2006ve,Du:2015wha}, and the higher \pT region in which other processes are expected to dominate, such as absorption due to colour-exchange interactions with the medium~\cite{Xu:2015bbz,Ferreiro:2012rq,Kopeliovich:2014una,PhysRevC.98.025207}. The naive expectation in this range is that recombination processes will partially couple the produced \jpsi to the hydrodynamic flow of the hot medium, resulting in an enhancement of the observed azimuthal asymmetry at lower \pT relative to higher \pT~\cite{Yan:2006ve,Kopeliovich:2014una}. In this picture, a flavour-dependent enhancement of the azimuthal asymmetry of \jpsi at low \pT can be interpreted as a difference in the degree of recombination between $c$- and $b$-quarks and it is expected that any flavour dependence will vanish at higher values of \pT, which are accessible by this measurement. A recent transport model study suggests a sensitivity of charm quarks to hydrodynamic flow~\cite{Li:2018leh}. In this model, additionally, a strong suppression of the prompt \jpsi yield in the final-state medium should lead to an azimuthal asymmetry even in the high \pT region~\cite{Kopeliovich:2014una}.

In non-central collisions, the overlap region of the colliding ions has an elliptic shape. The particle yield is influenced by this matter distribution, leading to the observation of an azimuthal anisotropy relative to the reaction plane as observed for charged hadrons~\cite{HION-2011-01,HION-2012-06,CMS-HIN-11-012,ALICE_Abelev:2012di,ALICE_Adam:2016izf}. The azimuthal distribution of particles is characterised by a Fourier expansion of the particle yield:
\begin{linenomath}
\begin{equation*}
\frac{\mathrm{d}N}{\mathrm{d}\phi} \propto 1 + \sum_{n=1}^{\infty} 2v_{n}\cos[n(\phi - \Psi_{n})]\text{,}
\end{equation*}
\end{linenomath}
where $\phi$ is the azimuthal angle of the particle relative to the detector frame of reference, and $\Psi_{n}$ is the $n$-th harmonic of the event-plane angle, which can be estimated using the event-plane method~\cite{Poskanzer:1998yz}. The second-order coefficient, $v_2$, is referred to as elliptic flow and its magnitude quantifies the yield modulation relative to the elliptical shape of the initial matter distribution.

Interestingly the observed azimuthal asymmetry for prompt \jpsi is the same in central collisions as in non-central  collisions~\cite{CMS2017}, although for inclusive \jpsi some indication of the centrality dependence is reported in the lower \pT region at forward rapidities at 5.02~\TeV~\cite{ALICE2017} and at 2.76~\TeV~\cite{ALICE2013}. This is in contradiction with the expected hydrodynamic behaviour, which is confirmed by the results for charged hadrons where the anisotropies are more significant in semi-central collisions than in peripheral and central collisions~\cite{HION-2011-01,HION-2012-06,CMS-HIN-11-012,ALICE_Abelev:2012di,ALICE_Adam:2016izf}. This intriguing observation may ultimately provide more insight into the origins of azimuthal asymmetries beyond a simple hydrodynamic picture. Further, there is evidence of a surprising universality among many different probes, such as $D$-mesons and jets~\cite{HION-2012-12,Sirunyan:2017plt}, for which the $v_2$ values are very similar at high \pT. This paper provides $v_2$ measurements as a function of transverse momentum, rapidity and collision centrality for both prompt and non-prompt \jpsi in the dimuon decay channel, extending the kinematic range covered by recent results from other LHC experiments~\cite{ALICE2013,ALICE2017,CMS2017}.
\section{ATLAS detector}
\label{sec:detector}
\newcommand{\AtlasCoordFootnote}{%
ATLAS uses a right-handed coordinate system with its origin at the nominal interaction point in the centre of the detector and the $z$-axis along the beam pipe. The $x$-axis points from the interaction point to the centre of the LHC ring, and the $y$-axis points upwards. Cylindrical coordinates $(r,\phi)$ are used in the transverse plane, $\phi$ being the azimuthal angle around the $z$-axis. The pseudorapidity is defined in terms of the polar angle $\theta$ as $\eta = -\ln \tan(\theta/2)$.}
The ATLAS detector~\cite{PERF-2007-01} at the LHC covers nearly the entire solid angle around the collision point. It consists of an inner tracking detector (ID) surrounded by a thin superconducting solenoid, electromagnetic and hadronic calorimeters, and a muon spectrometer (MS) incorporating three large superconducting toroid magnets.\footnote{\AtlasCoordFootnote}

A high-granularity silicon pixel detector surrounds the interaction region and typically provides four measurements per track. It is followed by a silicon microstrip tracker, which provides around eight two-dimensional measurement points per track. These silicon detectors are complemented by a transition radiation tracker (TRT), which enables radially extended track reconstruction up to $|\eta| = 2.0$. The ID system is immersed in a \SI{2}{\tesla} axial magnetic field and provides charged-particle tracking in the range $|\eta| < 2.5$.

The calorimeter system covers the pseudorapidity range $|\eta| < 4.9$. Within the region $|\eta|< 3.2$, electromagnetic calorimetry is provided by barrel and endcap high-granularity lead/liquid-argon (LAr) electromagnetic calorimeters, with an additional thin LAr presampler covering $|\eta| < 1.8$ to correct for energy loss in material upstream of the calorimeters. Hadronic calorimetry is provided by a steel/scintillator-tile calorimeter, segmented into three barrel structures within $|\eta| < 1.7$, and two copper/LAr hadronic endcap calorimeters covering $1.5 < |\eta| < 3.2$. The high $|\eta|$ region, $3.2 < |\eta| < 4.9$, is covered by forward copper/LAr and tungsten/LAr calorimeter (FCal) modules optimised for electromagnetic and hadronic measurements respectively.

The MS comprises separate trigger and high-precision tracking chambers measuring the deflection of muons in a magnetic field generated by the superconducting air-core toroid magnets. The precision chamber system covers the region $|\eta| < 2.7$ with three layers of monitored drift tubes, complemented by cathode strip chambers in the forward region, where the background is the highest. The muon trigger system covers the range $|\eta| < 2.4$ with resistive plate chambers in the barrel, and thin gap chambers in the endcap regions.

In addition to the muon trigger, two triggers are used in Pb+Pb collisions to select minimum-bias events. These are based on the presence of a minimum amount of transverse energy, of at least 50 GeV, in all sections of the calorimeter system with $|\eta| < 3.2$ or, for events which do not meet this condition, on the presence of energy deposits in both zero-degree calorimeters, which are primarily sensitive to spectator neutrons in the region $|\eta| > 8.3$.

A two-level trigger system is used to select events~\cite{TRIG-2016-01}. The first-level trigger is implemented in hardware and uses a subset of detector information to reduce the event rate to a design value of at most \SI{100}{\kHz}. This is followed by a software-based high-level trigger which reduces the event rate to about \SI{1}{\kHz}.
\section{Analysis}
\label{sec:analysis}
\subsection{Data, event selection and centrality definition}
Data from \PbPb collisions at $\sqn=5.02~\TeV$ were recorded by the ATLAS experiment in 2015. Events were collected using a trigger requiring at least two muons, both with $\pT^\mu > 4~\GeV$. This muon triggered dataset has an integrated luminosity of 0.42 \inb. In the offline analysis, reconstructed muons are required to satisfy the \emph{tight} muon \emph{working point} ignoring the TRT requirements~\cite{PERF-2015-10}, have $\pT^\mu > 4~\GeV$, $|\eta| < 2.4$, and be matched to the muons reconstructed at the trigger level. In addition, muon pairs are required to have pair $\pT > 9~\GeV$, rapidity $|y| < 2$ and be in the invariant mass range $2.6 < m_{\mu\mu} < 3.5~\GeV$. In addition to the muon triggered event sample, a minimum-bias triggered event sample and Monte Carlo (MC) simulated event samples were used for studies of the detector performance. Prompt ($\pp\to\jpsi\to\mu\mu$) and non-prompt ($\pp\to\bbbar\to\jpsi\to\mu\mu$) samples of \jpsi were produced using \textsc{Pythia} 8.212~\cite{Sjostrand:2014zea} for event generation and \textsc{Photos}~\cite{Barberio:1990ms} for electromagnetic radiation corrections. The A14 set of tuned parameters~\cite{ATL-PHYS-PUB-2014-021} is used together with CTEQ6L1 parton distribution function set~\cite{Pumplin:2002vw}. The response of the ATLAS detector was simulated using \textsc{Geant4}~\cite{SOFT-2010-01,AGOSTINELLI2003250}. Simulated events are overlaid on a sample of minimum-bias \PbPb events produced with \textsc{HIJING}~\cite{Gyulassy:1994ew} to replicate the high-multiplicity environment of heavy-ion collisions.

To characterise the \PbPb collision geometry, events are classified into centrality intervals determined  by the summed transverse energy deposited in the FCal, $\sum \ET^{\mathrm{FCal}}$, in each event. Centrality intervals are defined according to successive percentiles of the $\sum \ET^{\mathrm{FCal}}$ distribution ordered from the most central (the highest $\sum\ET^{\mathrm{FCal}}$, the smallest impact parameter) to the most peripheral collisions (the lowest $\sum \ET^{\mathrm{FCal}}$, the largest impact parameter). The average number of nucleons participating in the collision, \avgNpart, is calculated using a Glauber model analysis of the $\sum\ET^{\mathrm{FCal}}$ distribution~\cite{Miller:2007ri, Alver:2008aq}. The centrality intervals used in this analysis are quoted together with \avgNpart in Table~\ref{tab:reso_and_npart}. Only events in the 0--60\% centrality interval are used. Events in the 60--80\% are disregarded from the study, since they represent only a small fraction of events, below $3\%$ of the full \jpsi sample, which is not significant for the measurement.

\subsection{Signal extraction and observable determination}
The \jpsi $v_2$, the second-order coefficient of the Fourier decomposition of the azimuthal angle distribution, is measured using the event-plane method~\cite{Poskanzer:1998yz}. The event-plane angle is estimated by its second-order harmonic, $\Psi_2$, using the distribution of transverse energy deposited in the forward calorimeters. Similar methods are described in detail for previous azimuthal anisotropy analyses of charged hadrons  with the ATLAS detector~\cite{HION-2011-01,HION-2012-03,HION-2012-06}. To reduce autocorrelations in the event-plane analysis, $v_2$ is measured by correlating \jpsi with positive (negative) pseudorapidity with the event-plane angle measured using the FCal in the negative (positive) $\eta$-region. The prompt and non-prompt \jpsi yields are obtained from two-dimensional fits of the reconstructed \jpsi invariant mass and pseudo-proper decay time distributions. The azimuthal distributions of the prompt and non-prompt yields are fitted simultaneously to obtain the elliptic flow coefficients.

The second-order harmonic of the event-plane angle is determined using measurements of transverse energy deposits in each FCal system positioned at $\eta>3.2$  and $\eta<-3.2$. The flow vector is $\pmb{q}_2=\sum_{i} w_i(\cos(2\phi_i)\pmb{\hat x}+\sin(2\phi_i)\pmb{\hat y})$, where $\phi_{i}$ is the azimuthal coordinate of the $i^\text{th}$ calorimeter tower, $w_{i}$  is a weight that equals the transverse energy deposited in the calorimeter tower, and the sum is over all the FCal towers. The FCal towers consist of calorimeter cells grouped into regions in $\Delta\eta\times\Delta\phi$ of $0.1\times0.1$. The flow vector is determined separately in the positive and negative rapidity regions. The event-plane angle is then calculated as $2\Psi_{2} = \tan^{-1}(\pmb{q}_{2}\cdot\pmb{{\hat y}}/\pmb{q}_{2}\cdot\pmb{{\hat x}})$.

To ensure the uniformity of the event-plane angle distribution, the raw flow vector is corrected by subtracting its mean value, obtained by averaging over all events in a given centrality interval, so that $\pmb {q}_2=\pmb{q}_2^\text{raw}-\langle\pmb{q}_2^\text{raw}\rangle$. Since the mean values $\langle\pmb{q}_2^\text{raw}\rangle$ are found to be independent of the collision centrality, the $\langle\pmb{q}_2^\text{raw}\rangle$ averaged over all analysed events (centrality 0--60\%) is used. The remaining modulations of the event-plane angle distribution are removed by including a shift $\delta\Psi_{2} = \sum_{k=1}^{k_{\text{max}}}(1/k)\left[A_2 \cos(2k\Psi_2) + B_2 \sin(2k\Psi_2)\right]$~\cite{Poskanzer:1998yz}. The calculated coefficients $A_2=-\langle \sin(2k\Psi_2) \rangle$ and $B_2=\langle \cos(2k\Psi_2) \rangle$ are found to be centrality dependent up to $k=2$, and the sum is performed up to a conservative choice of $k_\text{max} = 12$. After these recentring and flattening corrections the event-plane angle distribution follows a uniform distribution.

The event-plane resolution, $\mathcal{R}$, is determined using the two-sub-event method~\cite{Poskanzer:1998yz} and the minimum-bias event sample.  Values of $\Psi_2$ are determined on both sides of the detector and are used to calculate $\mathcal{R} = \sqrt{\langle\cos(2\Delta\Psi_2)\rangle}$, where $\Delta\Psi_2$ is the difference between the values of $\Psi_2$ computed using the FCal modules in the positive and negative $\eta$-region of the detector. The resulting event-plane resolution depends strongly on centrality: it is poorer at low and high $\sum\ET^{\mathrm{FCal}}$ and better at middle values of $\sum\ET^{\mathrm{FCal}}$. The resolution is calculated in minimum-bias events in very fine bins of transverse energy. The average value of the resolution in wider bins must account for the different $\sum\ET^{\mathrm{FCal}}$ distribution of the sample of events containing \jpsi candidates. Thus, the event-plane resolution is weighted by the number of \jpsi candidates in a given centrality interval relative to the number of minimum-bias events in the same interval. The values of $\mathcal{R}$ for the centrality intervals used in this analysis are shown in Table~\ref{tab:reso_and_npart}.

\begin{table}[b!]
\centering
\caption{The average number of participating nucleons, \avgNpart, and the event-plane resolution, $\mathcal{R}$, with their total uncertainties in each centrality interval.}
\begin{tabular}{rrc}
\toprule
Centrality  & \multicolumn{1}{r}{\avgNpart}         & $\mathcal{R}$\\
\midrule
0--20\%    & $311.4 \pm 2.6$    & $0.759\pm0.011$ \\
20--40\%   & $160.3 \pm 2.7$    & $0.871\pm0.004$ \\
40--60\%   & $70.5 \pm 2.2$     & $0.766\pm0.006$ \\
\midrule
0--60\%    & $135.6 \pm 2.0$    & $0.794\pm0.032$ \\
\bottomrule
\end{tabular}
\label{tab:reso_and_npart}
\end{table}
To account for detector effects, each muon pair is corrected for trigger efficiency, $\epsilon_\text{trig}$, reconstruction efficiency, $\epsilon_\text{reco}$, and detector acceptance, $A$. These three quantities form a per-dimuon weight:
\begin{linenomath}
\begin{equation*}
w^{-1} = A\times\epsilon_\text{trig}\times\epsilon_\text{reco}\text{.}
\end{equation*}
\end{linenomath}
Trigger and reconstruction efficiencies are studied using the tag-and-probe method in data and in MC simulations as a function of the muon $\pT^\mu$ and $\eta^\mu$. The reconstruction efficiency increases from low to high $\pT^\mu$ and decreases from central to forward pseudorapidity, becoming constant at $\pT^\mu > 6~\GeV$ with a maximum efficiency of about 90\%.  Trigger efficiency increases from low to high $\pT^\mu$ and from central to forward pseudorapidity, increasing from 50\% to 85\% between the lowest and highest $\pT^\mu$. The acceptance is studied from MC simulations. It is defined as the probability that the \jpsi decay products fall within the fiducial volume $\pT^\mu > 4~\GeV$ and $|\eta^\mu|$ < 2.4 and assuming unpolarised \jpsi production~\cite{LHCb:Polarization,Alice:Polarization,,CMS:Polarization,}. A detailed description of the performance studies is presented in Ref.~\cite{Tapia:2018quy}.

The separation of the prompt and non-prompt \jpsi signals is performed using the pseudo-proper decay time of the \jpsi candidate, $\tau_{\mu\mu} = L_{xy}m_{\jpsi}/\pT$, where $L_{xy}$ is the distance between the position of the dimuon vertex and the primary vertex projected onto the transverse plane, $m_{\jpsi} = 3.096~\GeV$ is the value of the \jpsi mass~\cite{Patrignani:2016xqp}, and \pT is the transverse momentum of the dimuon system.

The corrected two-dimensional distribution of the number of events as a function of pseudo-proper decay time and dimuon invariant mass is used to determine the prompt and non-prompt \jpsi yields. The probability distribution function (PDF) for the fit is defined as a sum of five terms, where each term is the product of functions that depend on the dimuon invariant mass or pseudo-proper decay time. The PDF is written in a compact form as:
\begin{linenomath}
\begin{equation*}
    P(m_{\mu\mu},\tau_{\mu\mu}) = \sum^5_{i=1}N_i f_i(m_{\mu\mu}) \cdot h_i(\tau_{\mu\mu}) \otimes g(\tau_{\mu\mu}),
\label{eq:pdf}
\end{equation*}
\end{linenomath}
where $N_i$ is the normalisation factor of each component, $f_i(m_{\mu\mu})$ and $h_i(\tau_{\mu\mu})$ are distribution functions of the invariant mass, $m_{\mu\mu}$, and the pseudo-proper decay time, $\tau_{\mu\mu}$, respectively; $g(\tau_{\mu\mu})$ is the resolution function described by a double Gaussian distribution, and the $\otimes$-symbol denotes a convolution. The PDF terms are defined by Crystal Ball (CB)~\cite{Oreglia:1980cs}, Gaussian (G), Dirac delta ($\delta$), and exponential (E) distributions as specified in Table~\ref{tab:pdf}.
\begin{table}[t!]
\centering
\caption{Individual components of the probability distribution function in the default fit model used to extract the prompt and non-prompt contribution for $\jpsi$ signal and background. $F_\mathrm{CB}$ and $F_\mathrm{G}$ are the Crystal Ball (CB) and Gaussian (G) distribution functions respectively, $\omega$ is the relative fraction of the CB and G terms, $F_\mathrm{E}$ is an exponential (E) function, and $\delta(\tau_{\mu\mu})$ is the Dirac delta function.}
\begin{tabular*}{\figwidth}{c @{\extracolsep{\fill}} cccc}
\toprule
$i$ & Type & Source & $f_i(m_{\mu\mu})$ & $h_i(\tau_{\mu\mu})$ \\
\midrule
1 & Signal      & Prompt        & $\omega F_{\mathrm{CB}}(m_{\mu\mu})+(1-\omega)F_{\mathrm{G}}(m_{\mu\mu})$ & $\delta(\tau_{\mu\mu})$ \\
2 & Signal      & Non-prompt    & $\omega F_{\mathrm{CB}}(m_{\mu\mu})+(1-\omega)F_{\mathrm{G}}(m_{\mu\mu})$     & $F_{\mathrm{E}_1}(\tau_{\mu\mu})$ \\
3 & Background  & Prompt        & $F_{\mathrm{E}_2}(m_{\mu\mu})$                                            & $\delta(\tau_{\mu\mu})$ \\
4 & Background  & Non-prompt    & $F_{\mathrm{E}_3}(m_{\mu\mu})$                                        & $F_{\mathrm{E}_4}(\tau_{\mu\mu})$ \\
5 & Background  & Non-prompt    & $F_{\mathrm{E}_5}(m_{\mu\mu})$                                        & $F_{\mathrm{E}_6}(|\tau_{\mu\mu}|)$ \\
\bottomrule
\end{tabular*}
\label{tab:pdf}
\end{table}

The signal invariant mass shapes are described by the sum of a CB function and a single Gaussian function with a common mean. The width term in the CB function is equal to the Gaussian standard deviation times a scaling term, fixed from MC simulation studies. The CB left-tail and height parameters are also fixed from MC studies and variations of the two parameters are considered as part of the fit model's systematic uncertainties. The relative fraction of the CB and Gaussian functions, $\omega$, is free in the fit. The prompt background contribution to the invariant mass spectrum follows a nearly flat distribution, and is modelled by an exponential function, denoted $F_{\mathrm{E}_2}(m_{\mu\mu})$ in Table \ref{tab:pdf}. The non-prompt contribution to the background requires two exponential functions, denoted $F_{\mathrm{E}_3}(m_{\mu\mu})$ and $F_{\mathrm{E}_5}(m_{\mu\mu})$ in Table \ref{tab:pdf}, respectively.

The pseudo-proper decay time of the prompt signal is modelled with a Dirac delta function, while the non-prompt signal is described by a single-sided exponential, denoted $F_{\mathrm{E}_1}(\tau_{\mu\mu})$ in Table \ref{tab:pdf}. The backgrounds are represented by the sum of one prompt component and two non-prompt components. The prompt background component is described by a Dirac delta function. One of the non-prompt background contributions is described by a single-sided decay model (for positive $\tau_{\mu\mu}$ only), and the other is described by a double-sided decay model, denoted $F_{\mathrm{E}_4}(\tau_{\mu\mu})$ and $F_{\mathrm{E}_6}(|\tau_{\mu\mu}|)$ in Table \ref{tab:pdf}, accounting for candidates of mis-reconstructed or non-coherent muon pairs resulting from other Drell--Yan muons and combinatorial background. A double Gaussian resolution function, $g(\tau_{\mu\mu})$, is used in convolution with the background and signal terms. These resolution functions have a fixed mean at $\tau_{\mu\mu} = 0$ and free widths.

The free parameters in the fit are the number of signal candidates, the number of background candidates, the non-prompt fraction of signal candidates, the non-prompt fraction of background candidates, the non-prompt fraction of mis-identified candidates, the mean and width of the \jpsi mass peak, the slopes of the exponential distribution functions, and the widths of the pseudo-proper decay time resolution functions.

\begin{figure}[t!]
\begin{center}
    \includegraphics[width=8cm]{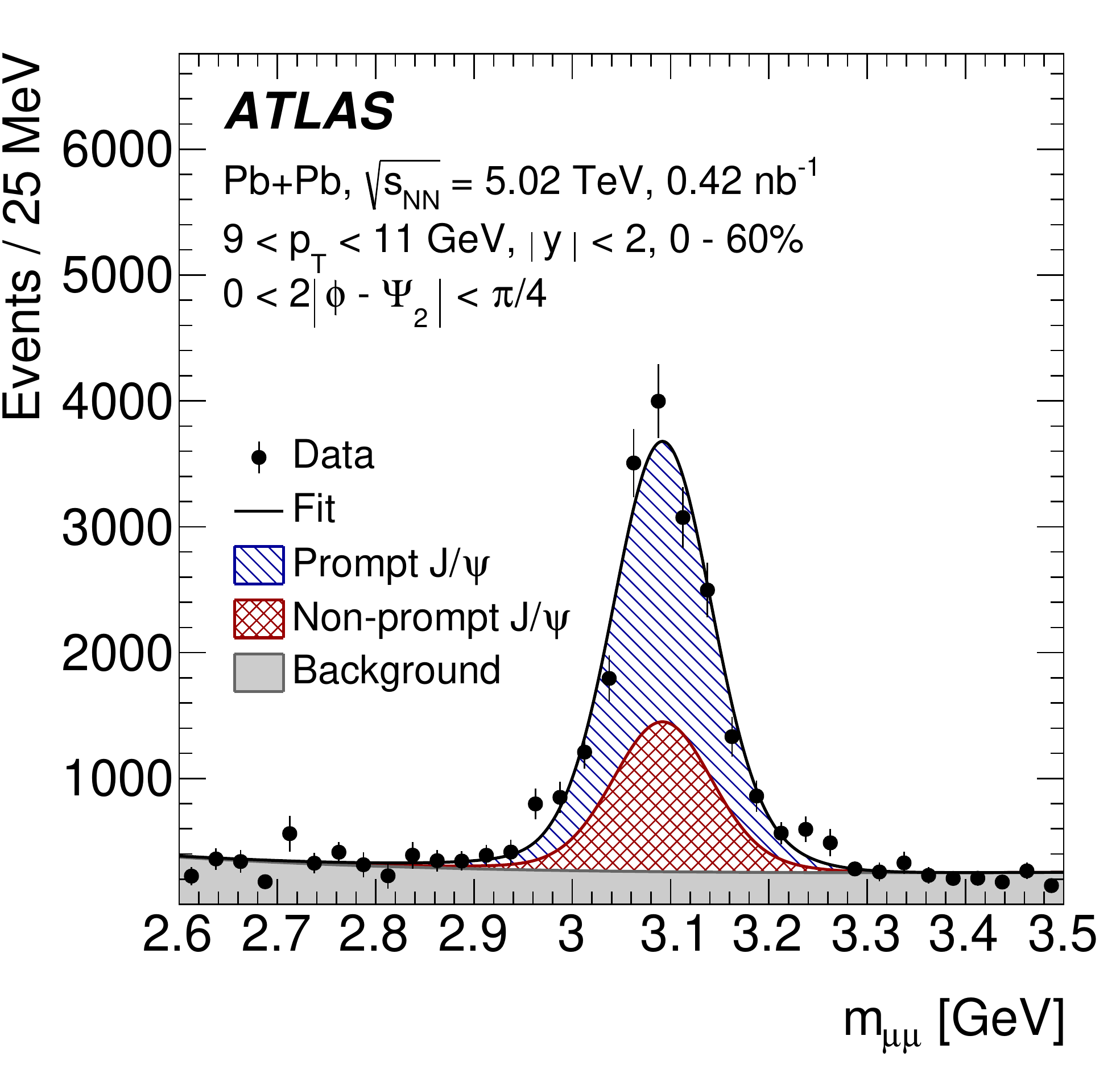}
    \includegraphics[width=8cm]{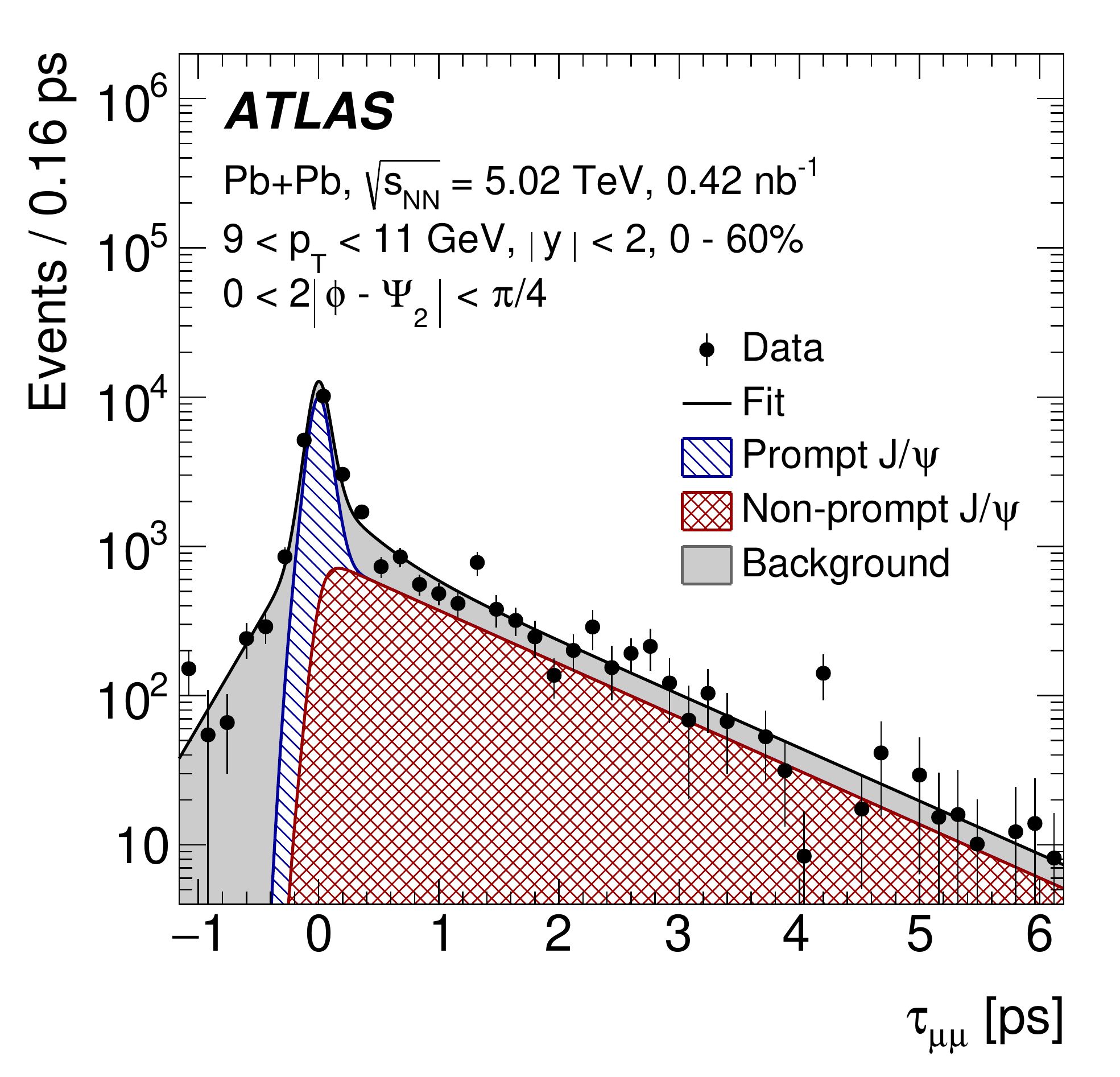}
\caption{Fit projections of the two-dimensional invariant mass ($m_{\mu\mu}$) and pseudo-proper decay time ($\tau_{\mu\mu}$) for the signal extraction for the azimuthal bin $0 < 2|\phi-\Psi_2| < \pi/4$ in the kinematic range $9 < \pT < 11~\GeV$, $0<|y|<2$ and 0--60\% centrality.}
\label{fig:mass_lifetime_fit}
\end{center}
\end{figure}

The relevant quantities extracted from the fit are: the number of signal candidates, $N_\text{signal}$, and the fraction of the signal that is non-prompt, $f_\text{NP}$. These are used to build azimuthal distributions of the prompt and non-prompt yields, as the fits are done in intervals of relative azimuthal angle $2|\phi-\Psi_2|$, $\pT$, $y$ and the collision centrality. Example plots of fit projections are shown in Figure~\ref{fig:mass_lifetime_fit}. The prompt and non-prompt signals are obtained from the fit as:
\begin{linenomath}
\begin{eqnarray*}
N_\text{prompt}     &=& N_\text{signal}(1 - f_\text{NP})\text{,} \\
N_\text{non-prompt} &=& N_\text{signal}f_\text{NP}\text{.}
\end{eqnarray*}
\end{linenomath}

While the total signal and the non-prompt fraction are weakly correlated, approximately less than 1\%, an artificial correlation is introduced when transforming these variables to the prompt and non-prompt yields. The sum of the two yields is constrained to the total number of signal candidates. To compute the correlation factor a toy Monte Carlo model is implemented using the same fit model and the output of the fits to data in bins of relative azimuthal angle, \pT, rapidity and centrality. The correlation varies with \pT from $-18\%$ to $-24\%$; in rapidity from $-22\%$ to $-16\%$; and is approximately constant as a function of centrality. The average correlation coefficient is $-20\%$ for all slices and azimuthal bins with a standard deviation of about 2\%. It is important  to note that this correlation is merely due to the procedure used to extract the yields from the invariant mass and pseudo-proper decay time distributions.

The elliptic flow coefficient is computed by fitting the prompt and non-prompt yields simultaneously to:
\begin{linenomath}
\begin{equation}
\frac{\mathrm{d} N}{\mathrm{d} (2|\phi-\Psi_2|)} = N_0 (1 + 2 v_2^{\text{fit}} \cos(2|\phi-\Psi_2|)),
\label{eq:model1}
\end{equation}
\end{linenomath}
in order to account for the anti-correlation between the two signals. This is achieved by minimising the $\chi^2$ function:
\begin{linenomath}
\begin{equation*}
\label{eq:chi211}
\chi^2(\pmb\theta) = (\pmb{y} - \pmb{\mu}(\pmb\theta))^{T} V^{-1} (\pmb{y} - \pmb\mu(\pmb\theta)),
\end{equation*}
\end{linenomath}
where $\pmb{y}$ is the vector of measurements, $\pmb{\mu}(\pmb\theta)$ is the vector of predicted values with parameters $\pmb\theta$, and $V$ is the error matrix. The two elements of the vector of measurements are the prompt and non-prompt yields; the vector of predicted values is given by Eq.~(\ref{eq:model1}) with the set of free parameters $\{N_{0},v_{2}^\text{fit}\}^\text{prompt}$ and $\{N_{0},v_{2}^\text{fit}\}^\text{non-prompt}$ for the modelling of the prompt and non-prompt yields respectively. The elements in the diagonal of $V$ are the yield uncertainties and the off-diagonal terms are the correlation terms between the prompt and non-prompt yields. 

\begin{figure}[b!]
\begin{center}
    \includegraphics[width=8cm]{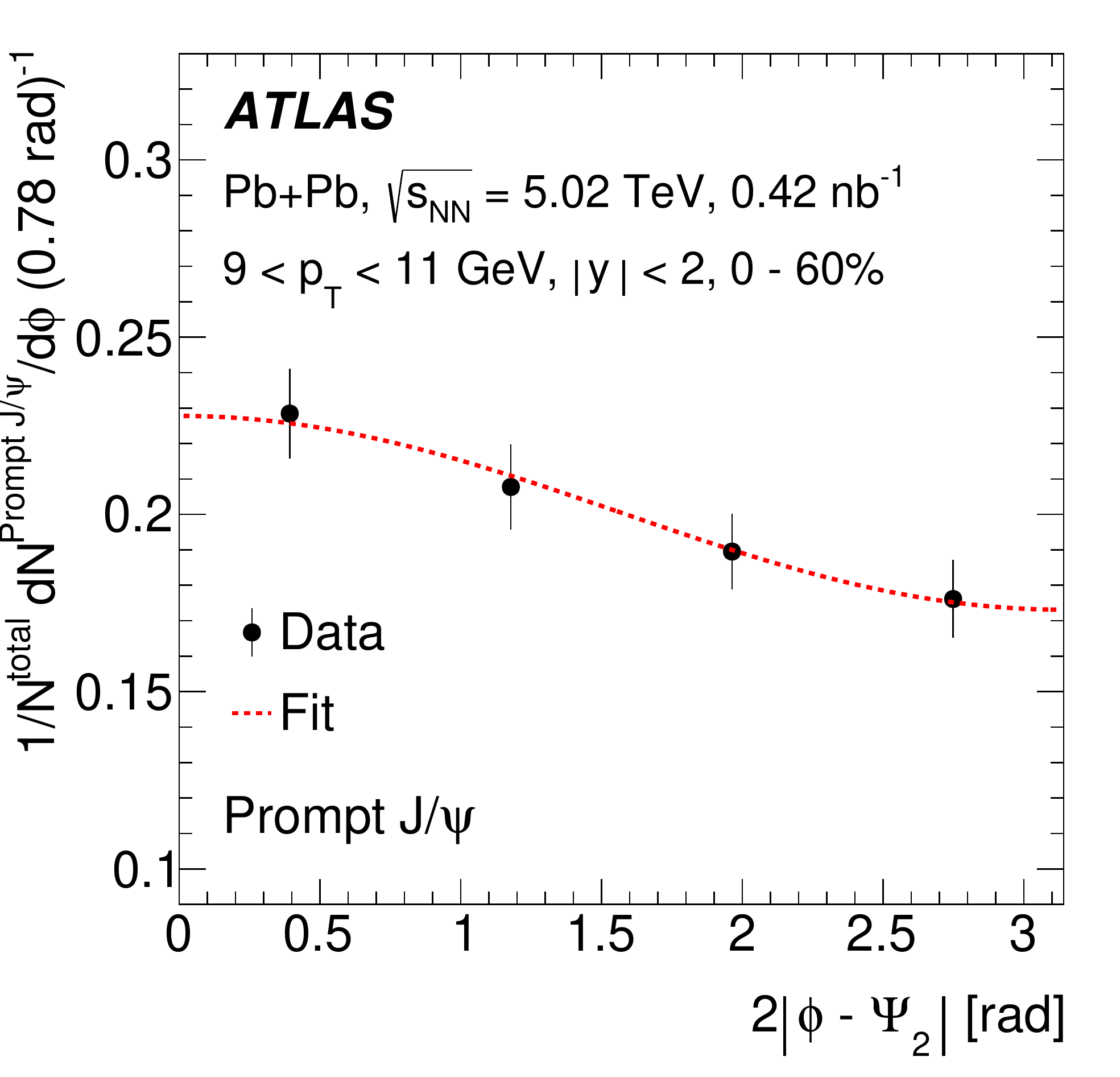}
    \includegraphics[width=8cm]{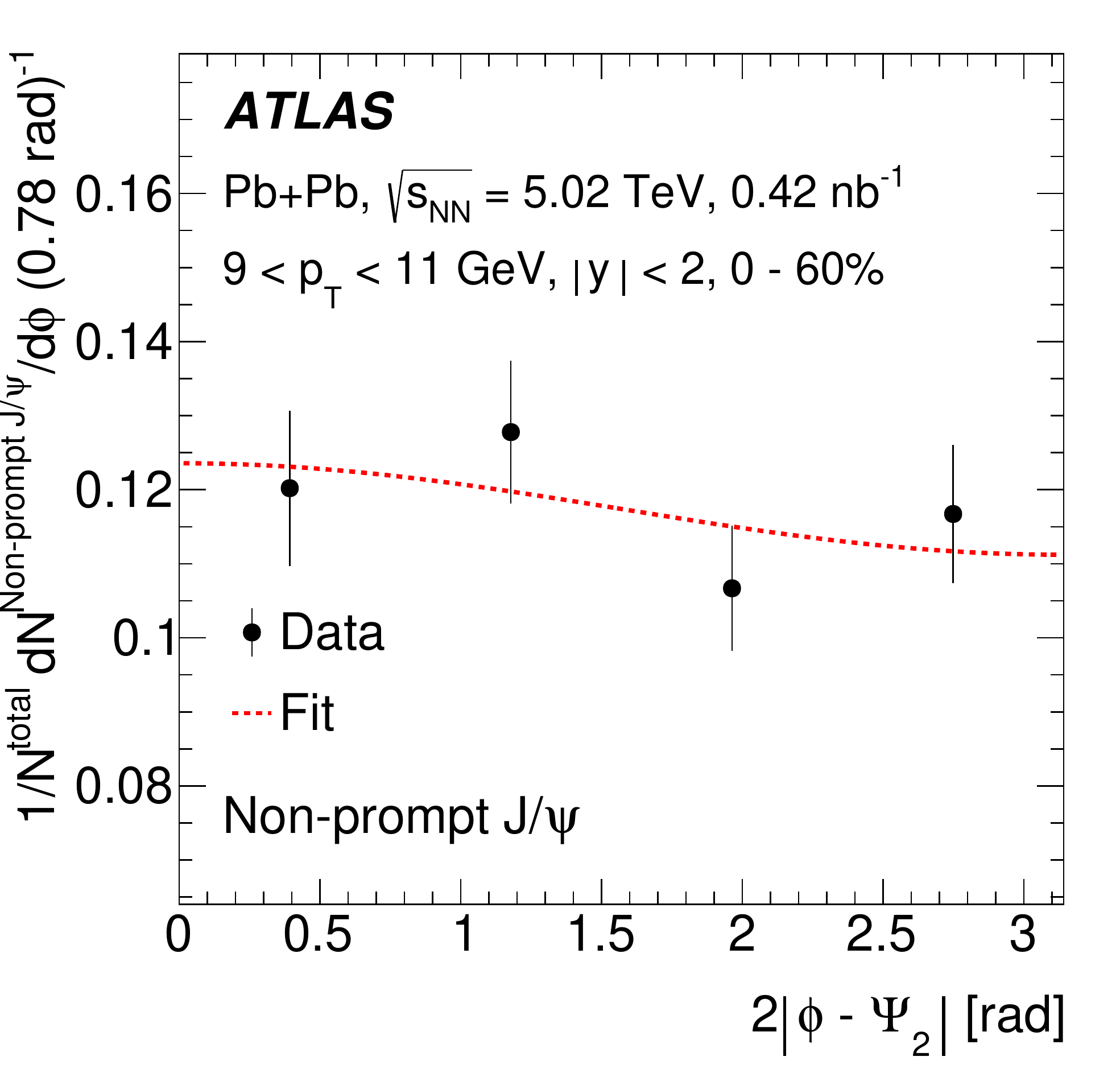}
\caption{The azimuthal distribution of prompt (left) and non-prompt (right) \jpsi yields for the lowest \pT bin studied. The yields are normalised by the inclusive \jpsi signal and the error bars are fit uncertainties associated with the signal extraction. The dotted red line is the result of the simultaneous fit used to compute $v_2$.}
\label{fig:flowfitresults}
\end{center}
\end{figure}

\begin{figure}[]
\begin{center}
    \includegraphics[width=8cm]{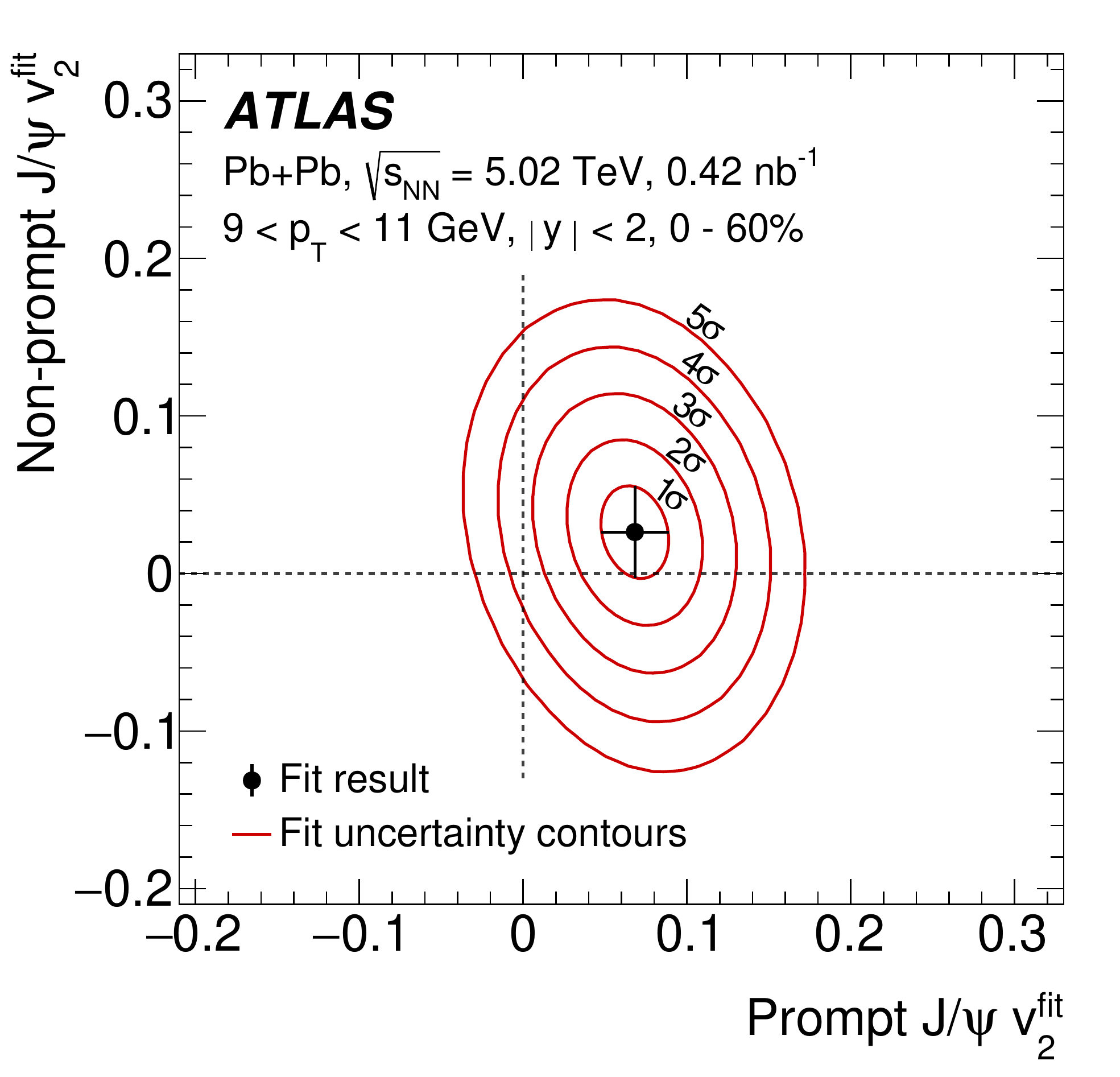}
\caption{Results of the error analysis for the fitted values of the prompt and non-prompt \jpsi $v_2$. The contour lines correspond to the $n\sigma$ fit uncertainties. For this bin, prompt \jpsi $v_2$ has a significance of $3\sigma$ and non-prompt \jpsi has a significance of $1\sigma$.}
\label{fig:flowfitcorrelation}
\end{center}
\end{figure}

An example of the prompt and non-prompt \jpsi yields normalised by the inclusive \jpsi yield and the projection of the fit result are shown in Figure~\ref{fig:flowfitresults}. The simultaneous fit of the prompt and non-prompt yields correctly accounts for the correlation between the two signals that arose from the modelling used for the signal extraction. The correlation between the fit parameters obtained from the simultaneous fit is shown in Figure~\ref{fig:flowfitcorrelation}.

In the final step, the fitted value of $v_{2}^\text{fit}$, is corrected for the event-plane resolution:
\begin{linenomath}
\begin{equation*}
v_2 = v_2^\text{fit}/\mathcal{R}.
\end{equation*}
\end{linenomath}
\subsection{Systematic uncertainties}
The systematic uncertainties of this measurement are classified into three groups: (a) related to the centrality definition, (b) related to the estimation of the event-plane method, and (c) related to extraction of the signal. The assigned systematic uncertainty from each source is defined in each bin of  \pT, rapidity or centrality as the root mean square of the difference between the nominal and varied values of the elliptic flow coefficient. All the uncertainties affecting the extraction of the signal are bin-to-bin uncorrelated, while the uncertainties related to the event-plane method can be correlated or uncorrelated, depending on the studied dependence.

The centrality intervals are defined by values of $\sum \ET^\text{FCal}$. These intervals have an uncertainty associated primarily to the effect of trigger and event selection inefficiencies as well as backgrounds in the most peripheral $\sum \ET^\text{FCal}$ intervals~\cite{HION-2012-06,ATLAS-CONF-2016-110}. To test the sensitivity of the results to this uncertainty, modified centrality intervals are used, for which the $\sum \ET^\text{FCal}$ cuts involved in the definition of the centrality intervals are shifted upward and downward, and the analysis is repeated. These changes affect the number of muon pairs entering the signal fitting procedure and thus have an impact on the final value of $v_2$. For the $v_2$ measurements as a function of \pT or rapidity the uncertainty is about $2\%$ for both prompt and non-prompt \jpsi $v_2$, while for the centrality dependence this source contributes a $10\%$ systematic uncertainty to both $v_2$ measurements.

For the estimation of the event-plane angle, the calibration coefficients for the recentring of the flow vector are calculated using a narrower centrality interval (20--60\%) instead of the full centrality range (0--60\%). For the evaluation of $\delta\Psi_2$ the sum limit is changed to $k_\text{max}=4$. No significant differences are observed, so a systematic uncertainty is not assigned. For the event-plane resolution, the three-sub-event method~\cite{Poskanzer:1998yz} is used as an alternative to compute $\mathcal{R}$ for the event-plane angle calculated with FCal in $\eta<0$ and $\eta>0$ independently. By using the electromagnetic and hadronic calorimeters, the event-plane angle is calibrated and determined in two different sections with $0.5<\eta<2$ and $-1.5<\eta<0$ and compared with the event-plane angle as measured in the FCal in $\eta<0$ to obtain its resolution. For the resolution of the FCal in the opposite side ($\eta>0$) a reflection of this selection is performed. Both the two-sub-event and three-sub-event  methods give consistent results for collisions in the 0--60\% centrality interval. To account for the different $\sum \ET^\text{FCal}$ distributions in minimum-bias and \jpsi triggered events, the difference between the resolutions computed in the two datasets is assigned as a systematic uncertainty and it is the dominant source of uncertainty. The total uncertainty for the average event-plane resolution adds a 4\% correlated uncertainty to the measurements integrated over centrality, while for the centrality dependence each point has an approximate $1.5\%$ uncertainty due to resolution. In the centrality interval considered in this analysis (0--60\%), it is found that the uncertainty related to the centrality definition has no effect on the event-plane resolution.

Many variations of the PDF defined for the signal extraction are considered. For the mass fit, the most important variations are the release of the fixed parameters of the CB function \cite{Oreglia:1980cs}, the substitution of the Gaussian + CB function by a single CB function, and variations of the Gaussian standard deviation and CB width scaling parameter. For the time dependence, the notable variations include that a single exponential function is replaced by short and long lifetime exponential decays, and one Gaussian function instead of two is considered for time resolution. Among all of these variations the biggest contribution is the release of the parameters of the CB function, which contributes between $10\%$ and $15\%$ uncertainty for the \pT and rapidity dependence of $v_2$ and up to $20\%$ for the centrality dependence. Deviations from the case of unpolarised \jpsi production are studied for different spin-alignment scenarios, corresponding to the extreme cases, as explained in Ref.~\cite{BPHY-2012-02}. These alternative scenarios are covered by a theoretical uncertainty of 3\% in $v_2$ for prompt \jpsi and 4\% for non-prompt \jpsi.

The choice of mass window used for the study is changed to analyse potential biases from the mass peak of the $\psi(2\mathrm{S})$. At high rapidity its width increases and the fit response for the background estimation changes. The mass window range is narrowed to $2.7 < m_{\mu\mu} < 3.4~\GeV$ and its impact is between $5\%$ and $10\%$ for the \pT, rapidity and centrality dependencies

The correlation between the prompt and non-prompt yields is also studied. It is either doubled or neglected, and shows a minor impact of $1\%$  for all presented results.
\section{Results}
\label{sec:result}
\begin{figure}[]
\begin{center}
    \includegraphics[width=8cm]{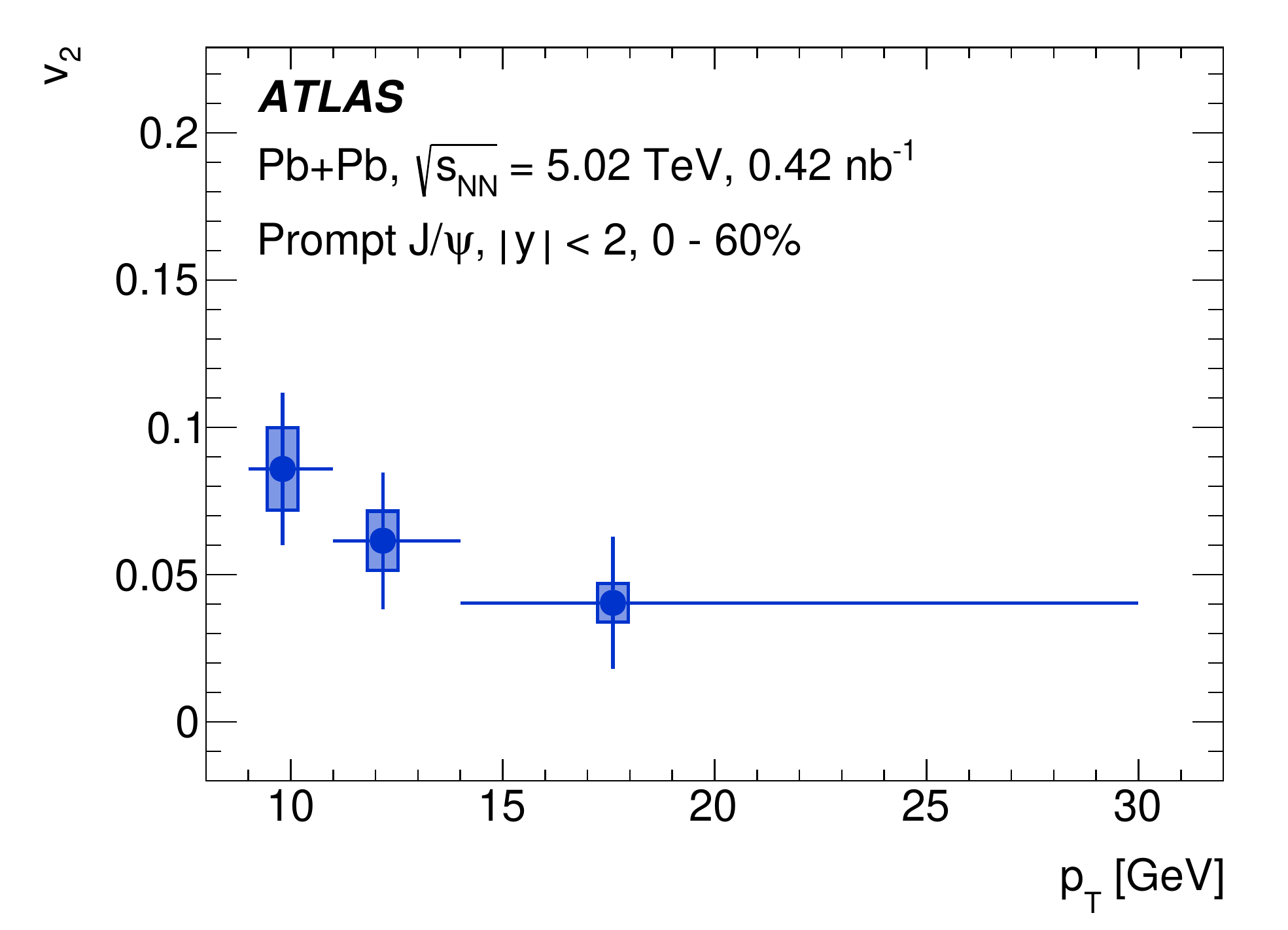}
    \includegraphics[width=8cm]{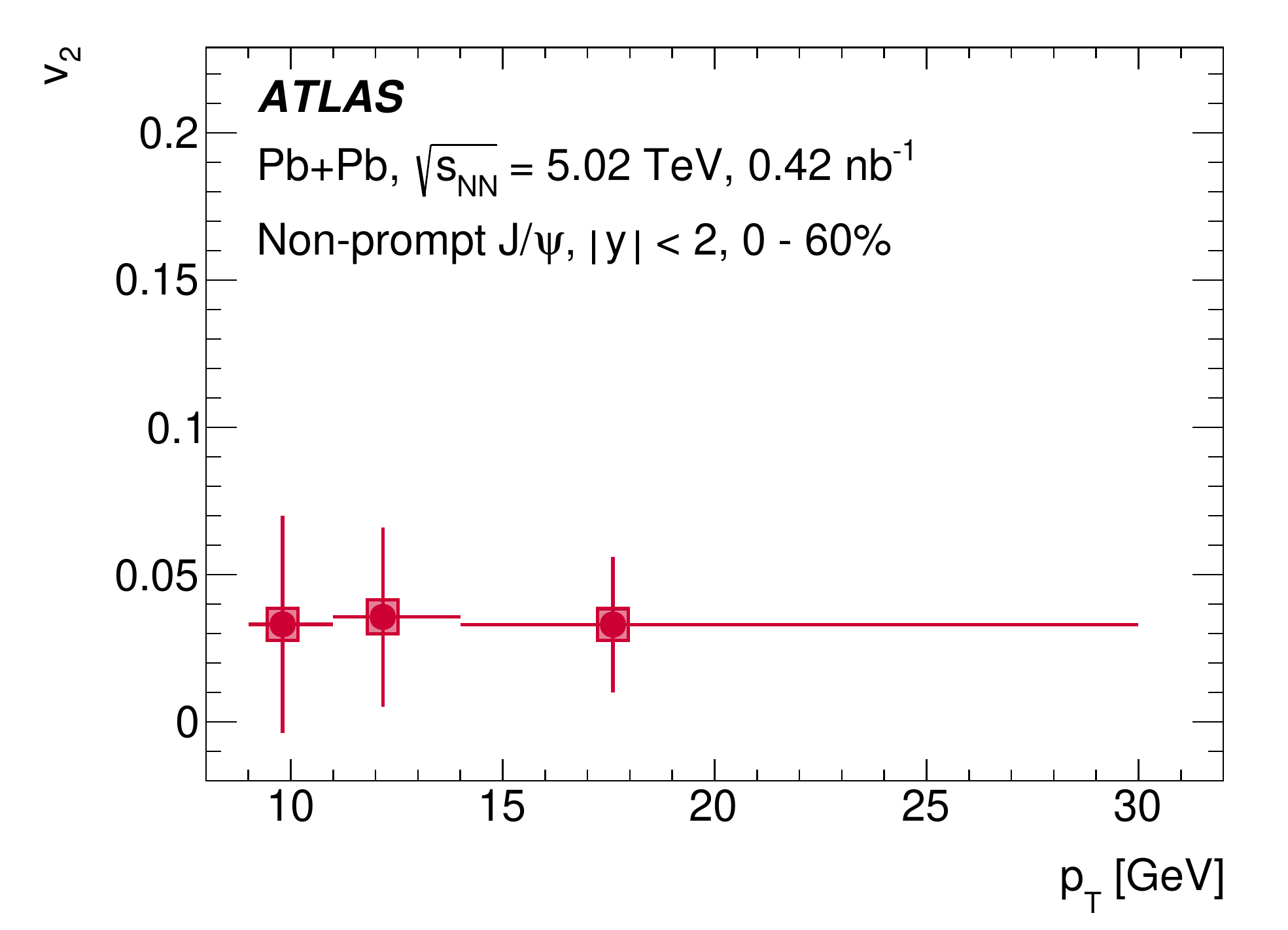}
\caption{Prompt (left) and non-prompt (right) \jpsi $v_2$ as a function of transverse momentum for the rapidity interval $|y| < 2$ and centrality 0--60\%. The statistical and systematic uncertainties are shown using vertical error bars and boxes respectively. The horizontal error bars represent the kinematic range of the measurement for each bin.}
\label{fig:v2pt}
\end{center}
\end{figure}

Results of $v_2$ measurements for prompt and non-prompt \jpsi are shown as a function of \pT in the range between 9 and 30~\GeV~in Figure~\ref{fig:v2pt}, for three \pT intervals. The centroid of each \pT bin is determined by the average \pT of the muon pairs in the corresponding bin and their values are $9.81$, $12.17$, and $17.6~\GeV$. The horizontal error bars correspond to the bin width reflecting the kinematic range of the measurement. The vertical error bars are the fit errors associated with statistical uncertainties, and the shaded boxes are the systematic uncertainties. The data are consistent with a non-zero flow signal in the full kinematic range studied ($9<\pT<30~\GeV$) for both prompt and non-prompt \jpsi.  As shown in Fig.~\ref{fig:flowfitcorrelation}, for the lowest \pT bin ($9 < \pT < 11~\GeV$), prompt \jpsi $v_2$ deviates from 0 with a significance of $3\sigma$ and non-prompt \jpsi with a significance of $1\sigma$. Prompt \jpsi exhibits a decreasing trend with a maximum value of $v_2$ close to 0.09 that decreases by nearly a factor of two over the whole studied kinematic range. The results for non-prompt \jpsi indicate a non-zero value with limited statistical significance. These $v_2$ values are consistent with being independent of \pT and compatible within uncertainties with the $v_2$ values of prompt \jpsi, particularly at the highest \pT.

\begin{figure}[]
\begin{center}
    \includegraphics[width=8cm]{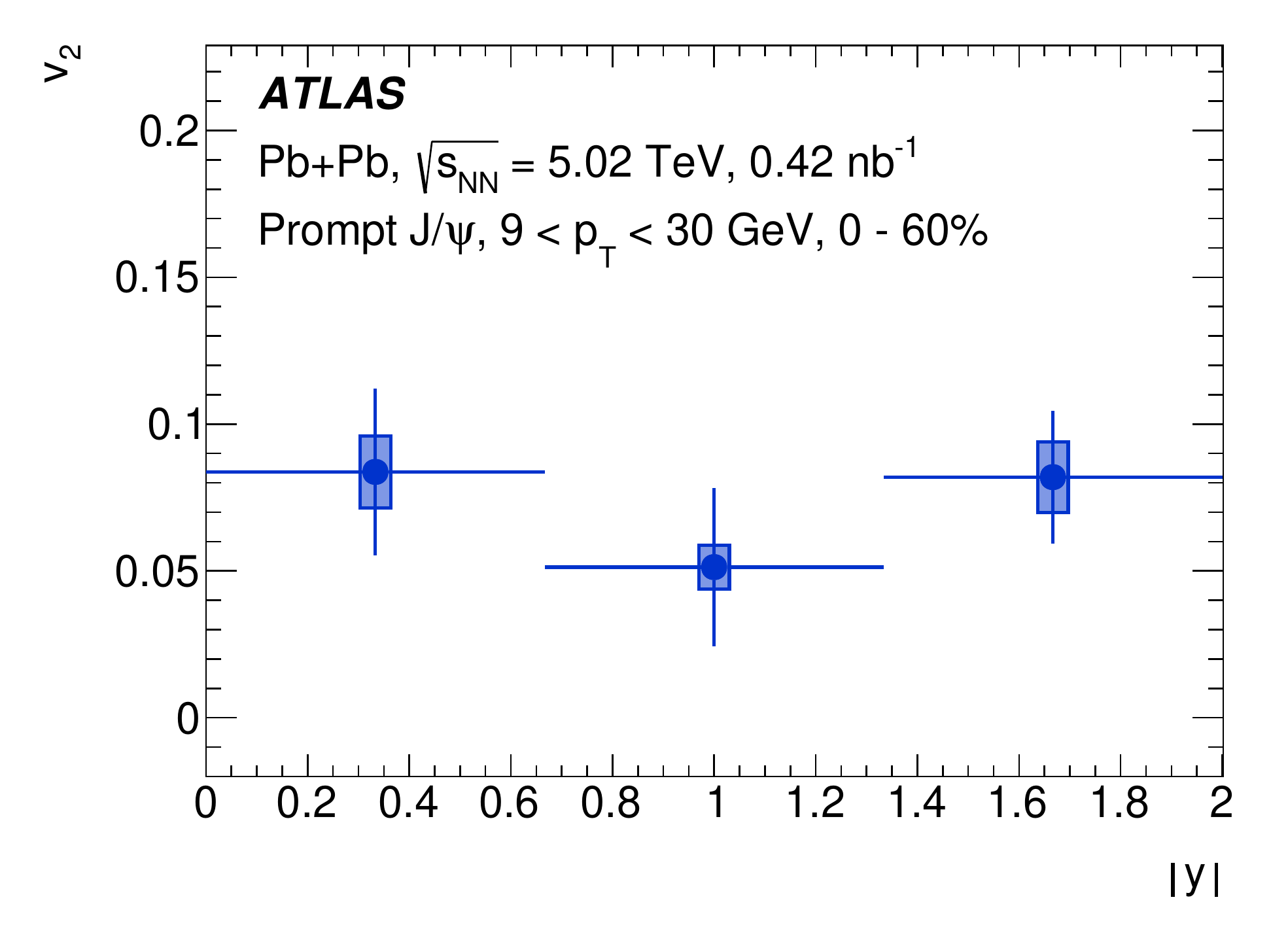}
    \includegraphics[width=8cm]{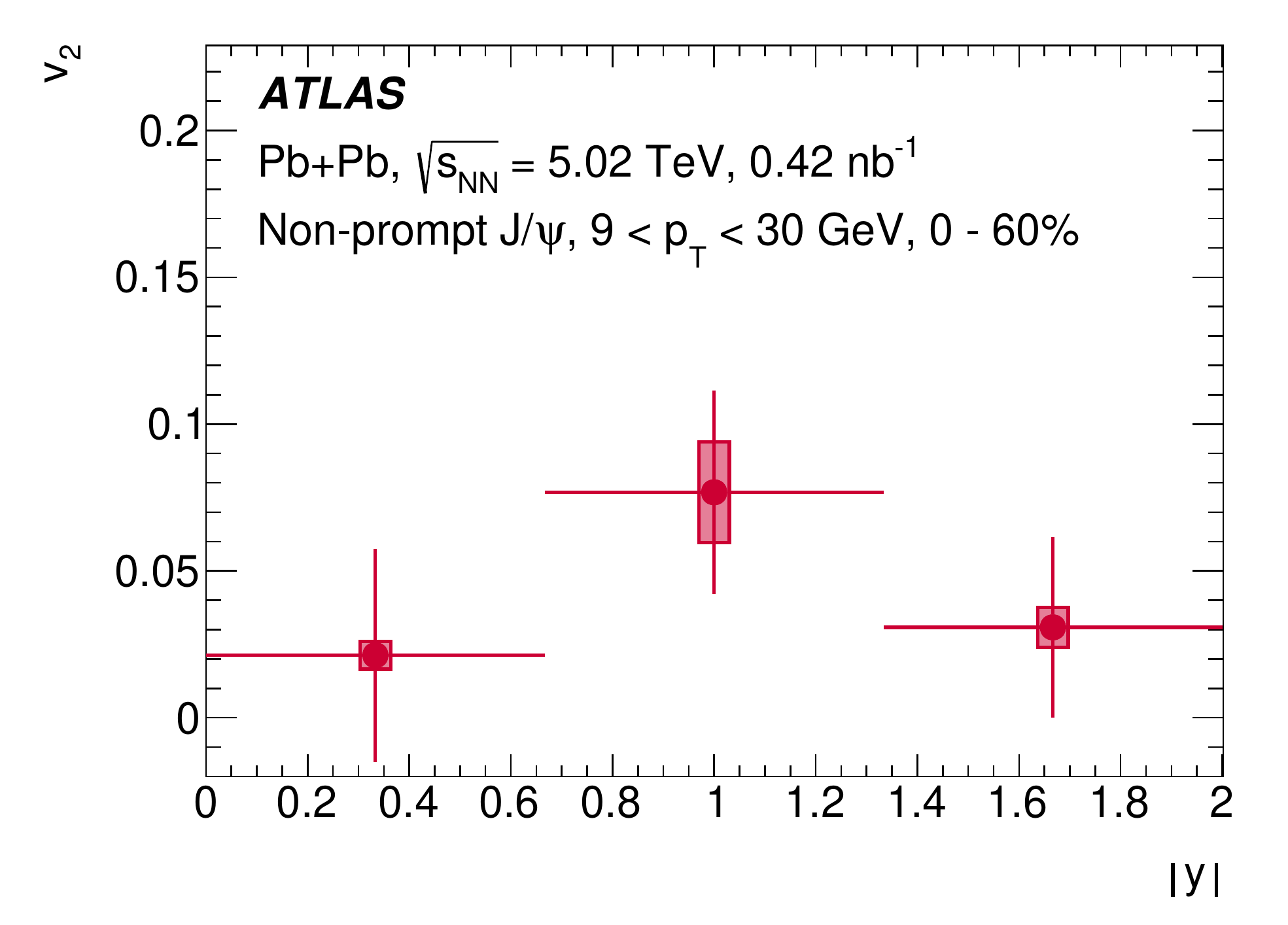}
\caption{Prompt (left) and non-prompt (right) \jpsi $v_2$ as a function of rapidity for transverse momentum in the range $9 < \pT < 30~\GeV$ and centrality 0--60\%. The statistical and systematic uncertainties are shown using vertical error bars and boxes respectively. The horizontal error bars represent the kinematic range of the measurement for each bin.}
\label{fig:v2Rap}
\end{center}
\end{figure}

\begin{figure}[]
\begin{center}
    \includegraphics[width=8cm]{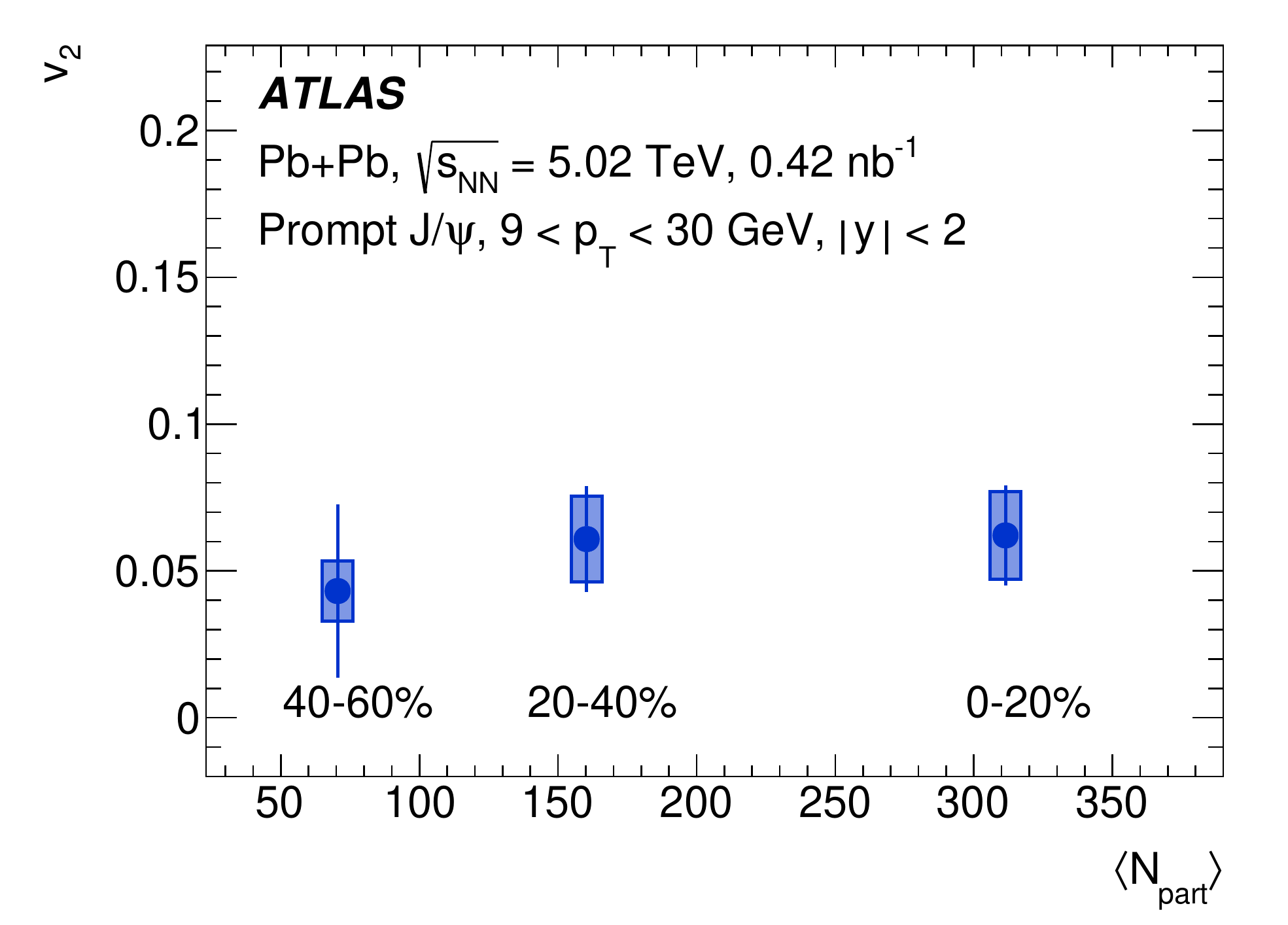}
    \includegraphics[width=8cm]{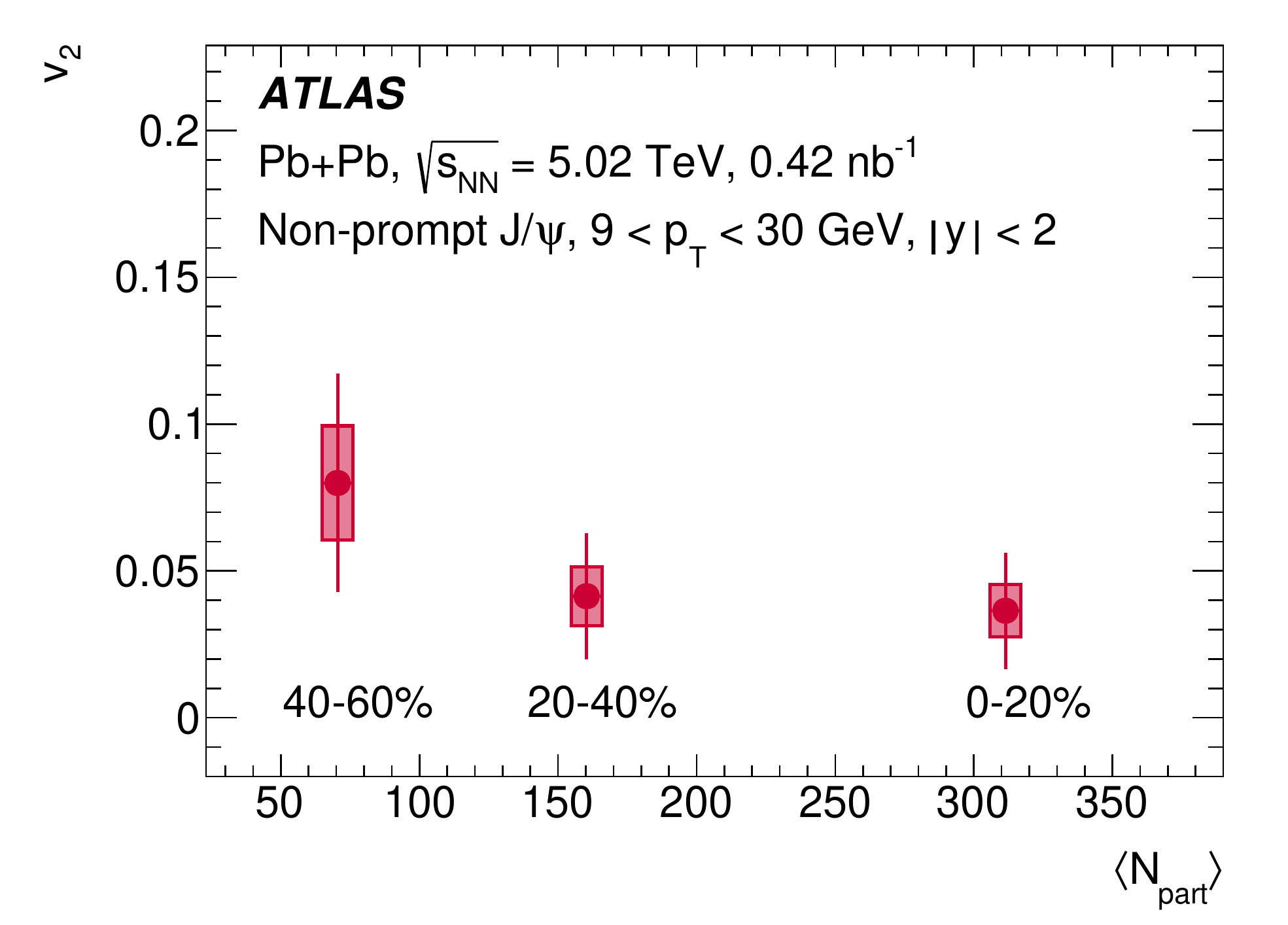}
\caption{Prompt (left) and non-prompt (right) \jpsi $v_2$ as a function of average number of nucleons participating in the collision for transverse momentum in the range $9 < \pT < 30~\GeV$ and rapidity $|y| < 2$. The statistical and systematic uncertainties are shown using vertical error bars and boxes respectively. The centrality interval associated to a given value of \avgNpart is written below each data point.}
\label{fig:v2Npart}
\end{center}
\end{figure}

The rapidity dependence of $v_2$ is shown in Figure~\ref{fig:v2Rap} and the centrality dependence in Figure~\ref{fig:v2Npart} for both prompt and non-prompt \jpsi. Neither shows significant rapidity or centrality dependence. The prompt \jpsi $v_2$ is larger than the non-prompt, in agreement with the larger values observed in the \pT dependence integrated over rapidity and centrality. The measured value  of $v_2$ for prompt \jpsi stays approximately the same in central collisions as in non-central collisions within uncertainties, in agreement with the observation of Ref.~\cite{CMS2017}. This is similar to the case of non-prompt \jpsi where no evident centrality dependence is observed within the uncertainties. This feature is in disagreement with the expected hydrodynamic behaviour  for charm quarks and may manifest a transition at medium \pT regime where there are thought to be different effects influencing \jpsi production~\cite{Zhou:2014kka,Xu:2015bbz,Yan:2006ve,Ferreiro:2012rq,Kopeliovich:2014una}.

\begin{figure}[]
\begin{center}
    \includegraphics[width=12cm]{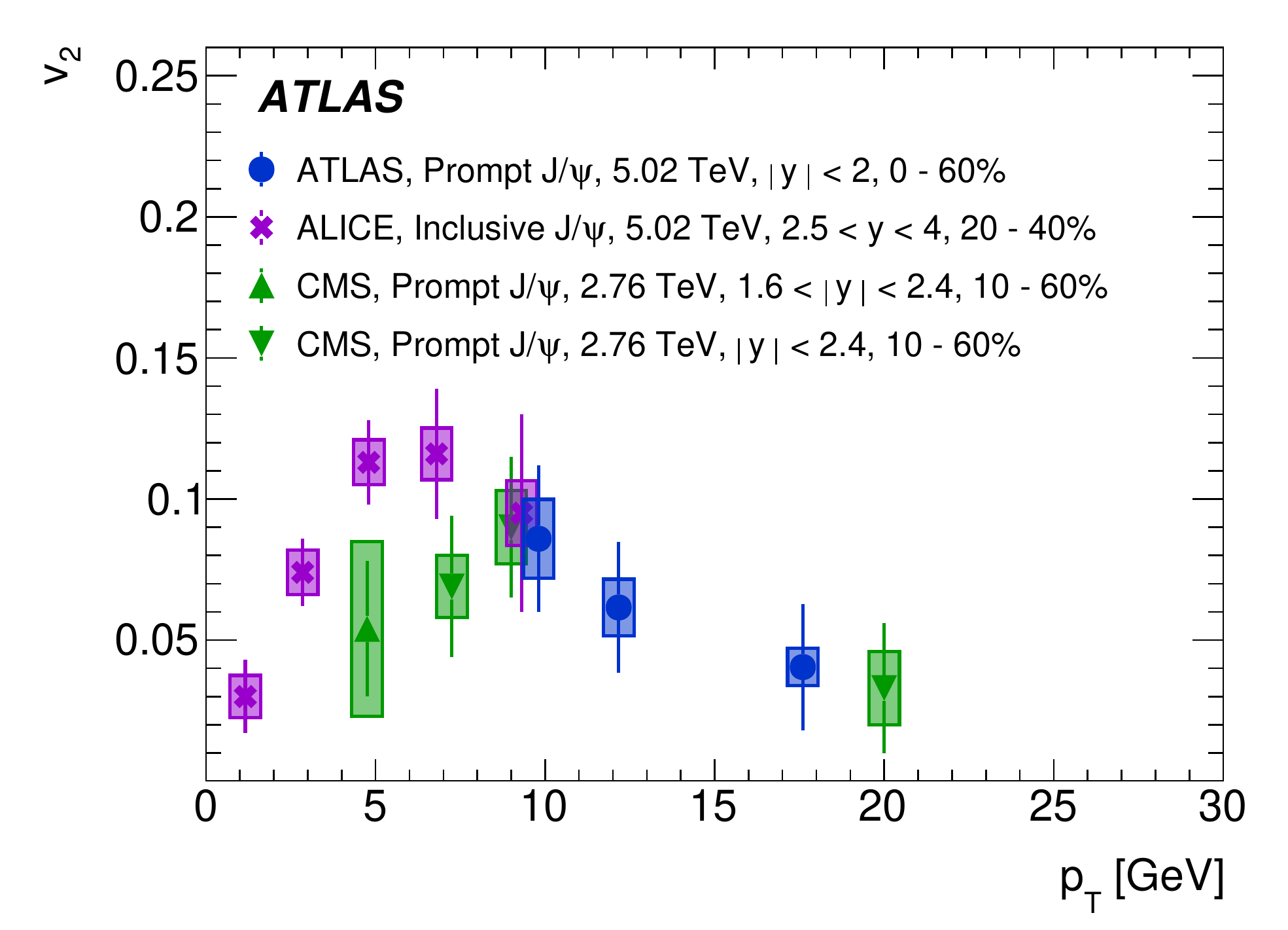}
\caption{Results for $v_2$ as a function of the transverse momentum of prompt \jpsi as measured by ATLAS in this analysis compared with inclusive \jpsi with $\pT < 12~\GeV$ as measured by ALICE at $5.02~\TeV$~\cite{ALICE2017}, and prompt \jpsi with \pT in the range $4 < \pT < 30~\GeV$ by CMS at $2.76~\TeV$~\cite{CMS2017}. The statistical and systematic uncertainties are shown using vertical error bars and boxes respectively.}
\label{fig:compare}
\end{center}
\end{figure}

In Figure~\ref{fig:compare} the available results for inclusive \jpsi ($\pT<12~\GeV$) from the ALICE experiment~\cite{ALICE2017} and prompt \jpsi (4 < \pT < 30~\GeV) from the CMS experiment~\cite{CMS2017} are compared with the results obtained in this analysis for prompt \jpsi ($9 < \pT < 30~\GeV$) as a function of the \jpsi transverse momentum. Despite different rapidity selections, the magnitudes of the elliptic flow coefficients are compatible with each other. Two features can be observed: first, the hydrodynamic peak is around 7 \GeV, a value that is significantly higher than what is observed for charged particles~\cite{HION-2011-01,HION-2012-06,CMS-HIN-11-012,ALICE_Abelev:2012di,ALICE_Adam:2016izf} where the peak is around 3 to 4 \GeV. This effect can be described qualitatively by thermalisation of charm quarks in the quark--gluon plasma medium with \jpsi regeneration playing a dominant role in the flow formation~\cite{Zhou:2014kka,Yan:2006ve}. The second feature is that $v_2$ has a substantial magnitude at high \pT. This can be connected with the suppression of \jpsi production due to mechanisms involving interactions with the medium such as absorption and melting~\cite{Kopeliovich:2014una} or energy loss~\cite{Arleo:2017ntr,Spousta:2016agr}.

\FloatBarrier
\section{Summary}
\label{sec:summary}
This paper presents measurements of the elliptic flow harmonic coefficients for \jpsi particles in the dimuon decay channel in $0.42~\mathrm{nb}^{-1}$ of \PbPb collisions recorded at $\sqn=5.02~\TeV$ with the ATLAS detector at the LHC. Results are presented for prompt and non-prompt \jpsi as a function of transverse momentum, rapidity and centrality. The measurement is performed in the \jpsi kinematic range $9<\pT<30~\GeV$, $|y|<2$, and 0--60\% centrality. The pseudo-proper decay time of the secondary vertex is used to separate the prompt and non-prompt components of \jpsi production and both yields are analysed simultaneously to properly assess the correlation between the two contributions.

A significant flow signal is found for prompt \jpsi, which decreases with increasing \pT. With limited statistical significance, it is found that non-prompt \jpsi $v_2$ is consistent with a flat behaviour over the studied \pT range. At high \pT, the prompt and non-prompt \jpsi $v_2$ values are compatible within the uncertainties. There is no evidence for a rapidity or centrality dependence for the prompt or non-prompt case. This suggests a similar underlying process describing the propagation of sufficiently high \pT charm and bottom quarks through the medium. The idea is supported by the recent observation of \jpsi yield suppression in \PbPb collisions by ATLAS, where a similar suppression pattern for prompt and non-prompt \jpsi is observed at high \pT. Additionally, this measurement covers the high \pT range of \jpsi and is found to be in a good agreement with previous reports, despite the different beam energy and rapidity selections.
\section*{Acknowledgements}

We thank CERN for the very successful operation of the LHC, as well as the
support staff from our institutions without whom ATLAS could not be
operated efficiently.

We acknowledge the support of ANPCyT, Argentina; YerPhI, Armenia; ARC, Australia; BMWFW and FWF, Austria; ANAS, Azerbaijan; SSTC, Belarus; CNPq and FAPESP, Brazil; NSERC, NRC and CFI, Canada; CERN; CONICYT, Chile; CAS, MOST and NSFC, China; COLCIENCIAS, Colombia; MSMT CR, MPO CR and VSC CR, Czech Republic; DNRF and DNSRC, Denmark; IN2P3-CNRS, CEA-DRF/IRFU, France; SRNSFG, Georgia; BMBF, HGF, and MPG, Germany; GSRT, Greece; RGC, Hong Kong SAR, China; ISF, I-CORE and Benoziyo Center, Israel; INFN, Italy; MEXT and JSPS, Japan; CNRST, Morocco; NWO, Netherlands; RCN, Norway; MNiSW and NCN, Poland; FCT, Portugal; MNE/IFA, Romania; MES of Russia and NRC KI, Russian Federation; JINR; MESTD, Serbia; MSSR, Slovakia; ARRS and MIZ\v{S}, Slovenia; DST/NRF, South Africa; MINECO, Spain; SRC and Wallenberg Foundation, Sweden; SERI, SNSF and Cantons of Bern and Geneva, Switzerland; MOST, Taiwan; TAEK, Turkey; STFC, United Kingdom; DOE and NSF, United States of America. In addition, individual groups and members have received support from BCKDF, the Canada Council, CANARIE, CRC, Compute Canada, FQRNT, and the Ontario Innovation Trust, Canada; EPLANET, ERC, ERDF, FP7, Horizon 2020 and Marie Sk{\l}odowska-Curie Actions, European Union; Investissements d'Avenir Labex and Idex, ANR, R{\'e}gion Auvergne and Fondation Partager le Savoir, France; DFG and AvH Foundation, Germany; Herakleitos, Thales and Aristeia programmes co-financed by EU-ESF and the Greek NSRF; BSF, GIF and Minerva, Israel; BRF, Norway; CERCA Programme Generalitat de Catalunya, Generalitat Valenciana, Spain; the Royal Society and Leverhulme Trust, United Kingdom.

The crucial computing support from all WLCG partners is acknowledged gratefully, in particular from CERN, the ATLAS Tier-1 facilities at TRIUMF (Canada), NDGF (Denmark, Norway, Sweden), CC-IN2P3 (France), KIT/GridKA (Germany), INFN-CNAF (Italy), NL-T1 (Netherlands), PIC (Spain), ASGC (Taiwan), RAL (UK) and BNL (USA), the Tier-2 facilities worldwide and large non-WLCG resource providers. Major contributors of computing resources are listed in Ref.~\cite{ATL-GEN-PUB-2016-002}.

\printbibliography
\clearpage 
 
\begin{flushleft}
{\Large The ATLAS Collaboration}

\bigskip

M.~Aaboud$^\textrm{\scriptsize 34d}$,    
G.~Aad$^\textrm{\scriptsize 99}$,    
B.~Abbott$^\textrm{\scriptsize 124}$,    
O.~Abdinov$^\textrm{\scriptsize 13,*}$,    
B.~Abeloos$^\textrm{\scriptsize 128}$,    
D.K.~Abhayasinghe$^\textrm{\scriptsize 91}$,    
S.H.~Abidi$^\textrm{\scriptsize 164}$,    
O.S.~AbouZeid$^\textrm{\scriptsize 39}$,    
N.L.~Abraham$^\textrm{\scriptsize 153}$,    
H.~Abramowicz$^\textrm{\scriptsize 158}$
% ...
\bigskip
\\

% ...
$^{13}$Institute of Physics, Azerbaijan Academy of Sciences, Baku; Azerbaijan.\\
% ...
$^{34}$$^{(a)}$Facult\'e des Sciences Ain Chock, R\'eseau Universitaire de Physique des Hautes Energies - Universit\'e Hassan II, Casablanca;$^{(b)}$Centre National de l'Energie des Sciences Techniques Nucleaires (CNESTEN), Rabat;$^{(c)}$Facult\'e des Sciences Semlalia, Universit\'e Cadi Ayyad, LPHEA-Marrakech;$^{(d)}$Facult\'e des Sciences, Universit\'e Mohamed Premier and LPTPM, Oujda;$^{(e)}$Facult\'e des sciences, Universit\'e Mohammed V, Rabat; Morocco.\\
% ...
$^{39}$Niels Bohr Institute, University of Copenhagen, Copenhagen; Denmark.\\
% ...
$^{91}$Department of Physics, Royal Holloway University of London, Egham; United Kingdom.\\
% ...
$^{99}$CPPM, Aix-Marseille Universit\'e, CNRS/IN2P3, Marseille; France.\\
% ...
$^{124}$Homer L. Dodge Department of Physics and Astronomy, University of Oklahoma, Norman OK; United States of America.\\
% ...
$^{128}$LAL, Universit\'e Paris-Sud, CNRS/IN2P3, Universit\'e Paris-Saclay, Orsay; France.\\
% ...
$^{153}$Department of Physics and Astronomy, University of Sussex, Brighton; United Kingdom.\\
% ...
$^{158}$Raymond and Beverly Sackler School of Physics and Astronomy, Tel Aviv University, Tel Aviv; Israel.\\
% ...
$^{164}$Department of Physics, University of Toronto, Toronto ON; Canada.\\
% ...
$^{a}$ Also at Borough of Manhattan Community College, City University of New York, NY; United States of America.\\
$^{b}$ Also at Centre for High Performance Computing, CSIR Campus, Rosebank, Cape Town; South Africa.\\
$^{c}$ Also at CERN, Geneva; Switzerland.\\
$^{d}$ Also at CPPM, Aix-Marseille Universit\'e, CNRS/IN2P3, Marseille; France.\\
$^{e}$ Also at D\'epartement de Physique Nucl\'eaire et Corpusculaire, Universit\'e de Gen\`eve, Gen\`eve; Switzerland.\\
% ...
$^{*}$ Deceased

\end{flushleft}


\end{document}